\documentclass[conference]{IEEEtran}
\IEEEoverridecommandlockouts
\usepackage{cite}
\usepackage{amsmath,amssymb,amsfonts}
\usepackage{algorithm}
\usepackage{algorithmic}
\usepackage{graphicx}
\usepackage{textcomp}
\usepackage{xcolor}
\usepackage{subfigure}
\usepackage{amsthm}
\newtheorem{definition}{Definition}

\def\BibTeX{{\rm B\kern-.05em{\sc i\kern-.025em b}\kern-.08em
    T\kern-.1667em\lower.7ex\hbox{E}\kern-.125emX}}
\begin{document}

\title{\texttt{STAD}: Spatio-Temporal Adjustment of Traffic-Oblivious Travel-Time Estimation}

\author{
\IEEEauthorblockN{Sofiane Abbar}
\IEEEauthorblockA{\textit{Qatar Computation Research Institute} \\
\textit{HBKU}\\
Doha, Qatar \\
sabbar@hbku.edu.qa}
\and
\IEEEauthorblockN{Rade Stanojevic}
\IEEEauthorblockA{\textit{Qatar Computation Research Institute} \\
\textit{HBKU}\\
Doha, Qatar \\
rstanojevic@hbku.edu.qa}
\and
\IEEEauthorblockN{Mohamed Mokbel}
\IEEEauthorblockA{\textit{Qatar Computing Research Institute} \\
\textit{HBKU}\\
Doha, Qatar \\
mmokbel@hbku.edu.qa}
}

\maketitle

\begin{abstract}

Travel time estimation is an important component in modern transportation applications. The state of the art techniques for travel time estimation use GPS traces to learn the weights of a road network, often modeled as a directed graph, then apply Dijkstra-like algorithms to find shortest paths. Travel time is then computed as the sum of edge weights on the returned path. In order to enable time-dependency, existing systems compute multiple weighted graphs corresponding to different time windows. These graphs are often optimized offline before they are deployed into production routing engines, causing a serious engineering overhead. In this paper, we present STAD, a system that adjusts -- on the fly -- travel time estimates for any trip request expressed in the form of origin, destination, and departure time. STAD uses machine learning and sparse trips data to learn the imperfections of any basic routing engine, before it turns it into a full-fledged time-dependent system capable of adjusting travel times to real traffic conditions in a city. STAD leverages the spatio-temporal properties of traffic by combining spatial features such as departing and destination geographic zones with temporal features such as departing time and day to significantly improve the travel time estimates of the basic routing engine. Experiments on real trip datasets from Doha, New York City, and Porto show a reduction in median absolute errors of 14\% in the first two cities and 29\% in the latter. We also show that STAD performs better than different commercial and research baselines in all three cities. 

\end{abstract}


\section{Introduction}

Real-time estimation of travel time is in the heart of modern transportation systems, spanning applications such as ride-sharing \cite{ma2013t}, driver dispatching \cite{yuan2012t,liao2003real}, and fleet management \cite{cristian2019multi}. 
For example, taxi and ride-sharing businesses heavily rely on travel time estimates for several core functionalities including route optimization, fare estimation, surge calculation, and taxi dispatching. 
%
With the abundance of floating vehicle data in the form of trips and GPS trajectories, it became possible to accurately estimate the travel time of trips. Three main schemes have been devised: segment-based, path-based, and origin-destination based. In segment-based approaches, GPS data is used to compute travel times for individual road segments. The travel time of a path is nothing but the sum of weights of its constituent edges~\cite{wedge18}. Path-based approaches aim at estimating travel times for sub-paths instead of individual segments, which allows them to capture some important inter-link transitions such as delays at junctions~\cite{yang2018pace}. 
Finally, origin-destination (OD) based approaches aim at estimating travel times without computing paths at all~\cite{li2018multi}. Despite this variety, traditional segment-based techniques are still preferred in large production and commercial systems, including Google Maps, HERE Maps, Apple Maps, TomTom, and MapBox~\cite{apple2018}. The reason is that most shortest-paths (routing) algorithms are optimized to work on directed graphs with static edge weights (travel times).  

\begin{figure}[h]
\centering
\subfigure[OSRM]{\label{fig:osrm_imp}\includegraphics[width=0.49\columnwidth]{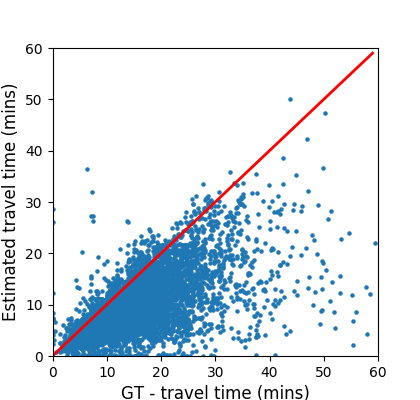}}
\subfigure[Google Maps]{\label{fig:google_imp}\includegraphics[width=0.49\columnwidth]{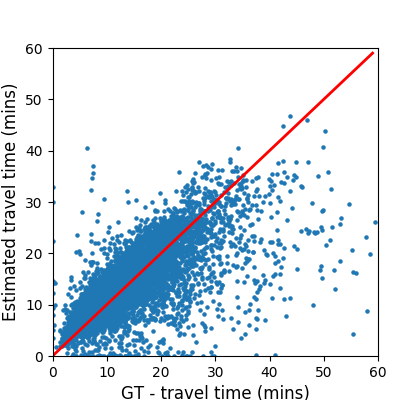}}
\caption{Scatter plots of actual trip travel time (x-axis) and estimated travel time (y-axis) by OSRM and Google Maps on the same set of 7750 taxi trips in Doha.} 
\label{fig:imperfections}      
\end{figure}

All routing engines that implement these techniques are imperfect to different degrees. Figure~\ref{fig:imperfections} illustrates these imperfections by plotting travel time estimates produced by two routing engines, namely OSRM~\cite{luxen-vetter-2011} (Fig~\ref{fig:osrm_imp}) and Google Maps (Fig~\ref{fig:osrm_imp}), compared to the actual trip duration reported for 7,750 taxi trips in Doha. Estimates where requested just before the trips started. Understandably, OSRM tends to underestimate travel times because it lacks traffic data. One way to fix this issue is to integrate time-dependency into OSRM, which requires the availability of large traffic data as well as a good knowledge on how to learn edge weights for different time windows, before one could deploy several instances of the routing engine that correspond to different time windows of interest (e.g., 24 hours of the day)~\cite{wedge18}. 
However, even the most sophisticated time-dependent routing engine can engender non negligible errors as shown in Figure~\ref{fig:google_imp}, in which we clearly see that even the premium Google Maps service that accounts for both historical and live traffic yields significant errors when it comes to estimating travel times.   

In this paper we describe a system, named \texttt{STAD}, that combines the best of segment-based and OD-based techniques to produce more accurate travel time estimates. 
\texttt{STAD} aims to quantify the spatio-temporal imperfections of any routing engine before it adjusts them to produce substantially more accurate travel time estimates. Besides, \texttt{STAD} can turn any static routing engine not designed for time-dependency into a full-fledged time-dependent engine with minimal sparse trips data and engineering overhead. Finally, \texttt{STAD} makes it easy for software developers, transport engineers, and researchers to develop in-house routing engines that can respond to live traffic data, whenever it is available to them.   

The main idea of \texttt{STAD} is to leverages well-known spatial and temporal characteristics of road traffic modeled as features in the machine learning module to learn personalized adjustments of travel times for any trip request $q_i$ in the format $<o_i, d_i, t_i>$ where $o_i$ and $d_i$ are respectively the origin and destination locations, and $t_i$ is the desired departure time. \texttt{STAD} consists of two components. The first one, is a routing engine, possibly traffic oblivious, which is optimized to generate free-flow travel time estimates (e.g., OSRM). The second component is a machine learning (ML) based module that mines sparse trip data to find out how to effectively adjust the travel time estimates of the base routing engine in a way that realistically reflects traffic dynamics related to the time of departure $t_i$ as well as $o_i$ and $d_i$ locations. 
Intuitively, what \texttt{STAD} aims to do is the following:\\ 
{\bf Example:} Assume that a user -- \textit{Leila} -- wants to go from home to work, the basic, free-flow, routing engine would return a decent path (route) with a travel time $\tau^{FF}=16mins$ and a total driving distance $l=12km$. However, looking the trip query, \texttt{STAD} would infer that {\it Leila} is departing at 07:15am, that she lives in {\it West Bay} which is in the city center, and works at {\it Education City} which is outside of city center. 
Historical trips data, would allow \texttt{STAD} to infer that {\it West Bay} is fairly congested during morning rush hours on weekdays, which allows it to adjust $\tau^{FF}$ with a $6mins$ offset, setting the travel time estimate of {\it Leila}'s trip request to $22mins$ instead. This is made possible by designing a set of features modeling the spatial and temporal traffic dependencies.

We also present \texttt{STAD$^*$}, a variant of \texttt{STAD} that is capable of integrating (near) live traffic data into the process of adjusting travel time estimates. Live traffic allows \texttt{STAD$^*$} to pick up  unexpected traffic events as accidents, road works, or any unusual congestion not seen in the historic trip data used to train the ML module. This is made possible by tracking a global and per zone congestion indices ($\mathcal{CI}$) computed as the ratio between the speeds achieved by recent trips (e.g., last 10 minutes) versus expected speeds learnt from historic trips data for the same time window.\\
{\bf Example (Cont'd):}
Going back to {\it Leila}'s example, assuming that historic trip data reveals an expected (mean) speed of $45km/h$ for the time window 07:00-08:00am. However, the recently completed trips -- in the previous $10mins$ prior to Leila's query (i.e., 07:05am - 07:15am) -- reported an average speed of $60km/h$ due school vacation. Then, \texttt{STAD}$^*$ can take this information into account and produce a lower estimated travel time of $19mins$ instead of $22mins$.    

The experimental evaluations we conducted using real taxi trip datasets from Doha (Qatar), NYC (US), and Porto (Portugal) demonstrate the effectiveness of \texttt{STAD} when compared to different baselines from industry (Google Maps) and research (\texttt{KNN}~\cite{wang2016simple,wang2019simple}). 
Indeed, our system achieves a median relative percentage error of 16.5\% compared to 18.55\% for \texttt{KNN} and 18.40\% for Google Maps in Doha. In terms of median absolute errors, \texttt{STAD}'s travel time estimation are $\pm 126$ seconds of ground truth travel times in Doha, which is 13\% more accurate than \texttt{KNN} ($\pm 144$ seconds) and 14\% better than Google Maps($\pm 146$ seconds). Similarly in other cities,  \texttt{STAD}'s travel time estimates are 15\% more accurate than \texttt{KNN} in NYC and 29\% in Porto. Experiments also show that live traffic data can improve the accuracy of travel times as the median absolute error of \texttt{STAD}$^*$ is 5 seconds lower than that of \texttt{STAD}. For transparency and reproduciblily of results, we voluntarily share on Github\footnote{https://github.com/vipyoung/stad} the source code of different algorithms and baselines developed along with datasets used.


\textit{Roadmap.} The remainder of the paper is organized as follows. Section~\ref{sec:architecture} depicts the general architecture of \texttt{STAD}.  Section~\ref{sec:problem} defines our general concepts and provides a formal definition of the travel time estimate adjustment problem. Section~\ref{sec:stad} introduces -- \texttt{STAD} -- a system capable of adjusting free-flow travel-time estimates. In Section~\ref{sec:stad_star} we present \texttt{STAD}$^*$, a variant of our system capable of integrating (near) live traffic data. Section~\ref{sec:experiments} presents the results of our experimental evaluation on three real traffic datasets. Section~\ref{sec:related} described the related work. Section~\ref{sec:conclusion} concludes the paper with some remarks and future directions.  


\section{Architecture}
\label{sec:architecture}
\begin{figure}[h]
\centering
\includegraphics[width=0.98\columnwidth]{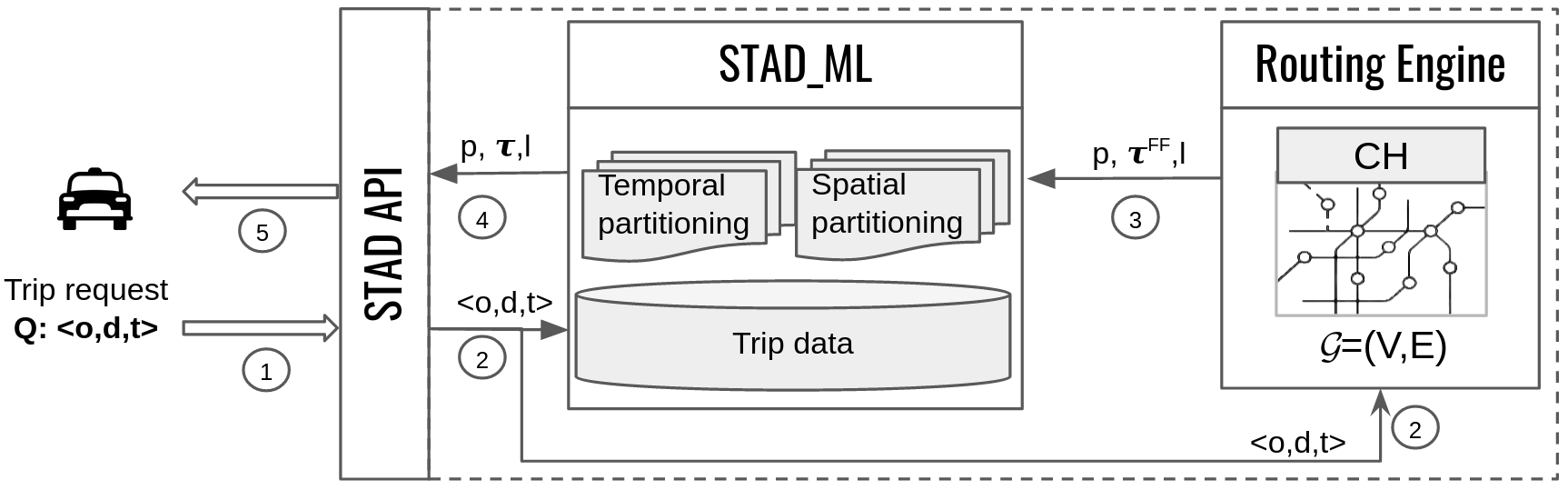}
\caption{High-level overview of STAD architecture.}
\label{fig:overview}
\end{figure}

The general architecture of \texttt{STAD} is depicted in Figure~\ref{fig:overview} which shows the two main components of the system: A base routing engine and a machine learning module for spatio-temporal adjustment of travel time estimates. 

The workflow of \texttt{STAD} is as follows. Upon the reception of a trip request $q=(o,d,t)$ where $o$ is the origin location, $d$ is the destination, and $t$ is the departure time, STAD API forwards the request to both components. 
At this stage, the machine learning module will simply extract some spatial and temporal features from the request itself. This includes for instance the geographic zones to which $o$ and $d$ belong, as well as the hour of the day and day of the week by decoding $t$.  
At the same time, the routing engine generates a triplet $(p, \tau^{FF}, l)$ where $p$ is a path between $o$ and $d$, $\tau^{FF}$ is the estimated travel time (free-flow), and $l$ is the path length (driving distance). The triplet is pushed to the machine learning module, which combines $\tau^{FF}$ and $l$ with the previously extracted spatial and temporal features to compute an accurate travel time $\tau$, which is an adjustment of $\tau^{FF}$, that is returned to the client along with other route related information.  

This architecture enables \texttt{STAD} to simply yet comprehensively support time-dependent travel time estimation and routing queries. It also makes it easy to digest live traffic data into these processes by compounding all these aspects into the machine learning module. 


\section{Background and Problem Formulation}
\label{sec:problem}

We define here after some basic concepts related to our proposal before we formalize our problem.

\begin{definition}{Road Network.}
A road network is represented as a directed weighted graph $\mathcal{G}=(V, E)$. $V$ is a set of nodes and $E$ is the set of directed weighted edges (road segments). Each edge $e_i$ comes with a length and possibly a set of intermediary points representing its geometry, a max speed value, and a type of road category. In this work, only length is considered. Each node $v_i \in V$ is associated with a pair of (latitude,longitude) coordinates to accurately position it on the map. 
\end{definition}

\begin{definition}{Path.}
A path is a sequence of connected consecutive edges $p=\{e_1, e_2, \ldots e_k\} | e_i\in E$. The length of a path $p_i$ is the sum of lengths of its edges, and is denoted $l_i$.

\end{definition}

\begin{definition}{Trajectory.}
A trajectory $tr_i=\{s_1^i, s_2^i, \dots s_n^i\}$ is a sequence of timestamped positional sample points ($s_j^i$) generated by GPS enabled devices of floating vehicles. Each point $s_j$ comes with spatial coordinates ($lon_j, lat_j$) and a timestamp $t_j$. Trajectories can be map-matched to $\mathcal{G}$ to find their corresponding paths. 
\end{definition}

\begin{definition}{Trip.}
A trip ($\Gamma$) is a tuple of the following attributes: $\Gamma_i=(o_i, d_i,t_i,\tau_i,tr^i,p_i)$. $o_i$ and $d_i$ denote the origin and destination locations respectively; $t_i$ and $\tau_i$ denote the departure time and the travel time respectively; $t_r^i$ and $p_i$ denote the trajectory and its corresponding path respectively if available. $\mathcal{T}=\{\Gamma_1, \ldots \Gamma_m\}$ is the set of trips available to us.
\end{definition}

\begin{definition}{Routing Engine}
A routing engine is an optimized system built on top of the road graph $\mathcal{G}$, capable of executing basic routing operations such as: navigation (for shortest paths), map-matching (for most likely path covering a trajectory), distance matrix (for tables of travel times and/or distances between sets of departures and destinations). We use hereafter OSRM~\cite{luxen-vetter-2011}, a highly regarded open source project that many commercial companies use for routing operations. 
\end{definition}





\textbf{Problem Definition.}
At a high level, our problem can be announced as follows: 
Given a road network graph $\mathcal{G}$, a set of trips data $\mathcal{T}$, and a trip request $q=<o, d, t^d>$ where $o=(lon^o,lat^o)$ and $d=(lon^d,lat^d)$ are origin and destination locations, and $t^d$ is the departure time, find the most likely travel time $\hat{\tau}$ for $q$.   
However, given the two stages nature of our system, we can reformulate the problem as follows: 
Given a basic routing engine, a set of trips $\mathcal{T}$, and a trip query $q$, first compute the free-flow travel time $\tau^{FF}$ and path length $l$ for $q$ using the basic routing engine; Next, learn to adjust $\tau^{FF}$ to reflect actual traffic patterns covered by $\mathcal{T}$ and produce more accurate travel time estimate $\hat{\tau}$. 

\section{STAD: Spatio-Temporal Adjustment of Travel Time Estimates}
\label{sec:stad}

In this section, we describe our system, \texttt{STAD}, for \underline{S}patio-\underline{T}emporal \underline{AD}justment of traffic oblivious travel times. We first give a brief overview of the system. Then, we explain how space and time are partitioned to capture varying spatio-temporal impact of traffic on travel time estimation. Next, we discus the features we used and the choice of the machine learning algorithm. Finally, we describe the online adjustment process of travel times.  

\subsection{General overview}

\texttt{STAD} consists of two main components: (i.) A base routing engine used to produce free-flow paths and travel time estimates between pairs of locations. One can think of this as simply running Dijkstra algorithm for shortest path on a directed weighted graph representing the road network, where weights are traversal times of edges derived from length and default speeds. 
(ii.) A machine learning module that uses spatial and temporal features to adjust free-flow travel time estimates to traffic patterns captured in the trips data. 
These components are first used offline to learn the best parameters possible for the adjustment of travel times in a given city, then online to adjust the travel time of any trip request. 

\subsection{Spatial and temporal partitioning}
\begin{figure}
\centering
\subfigure[Administrative zones]{\label{fig:sparse_doha}\includegraphics[width=0.49\columnwidth]{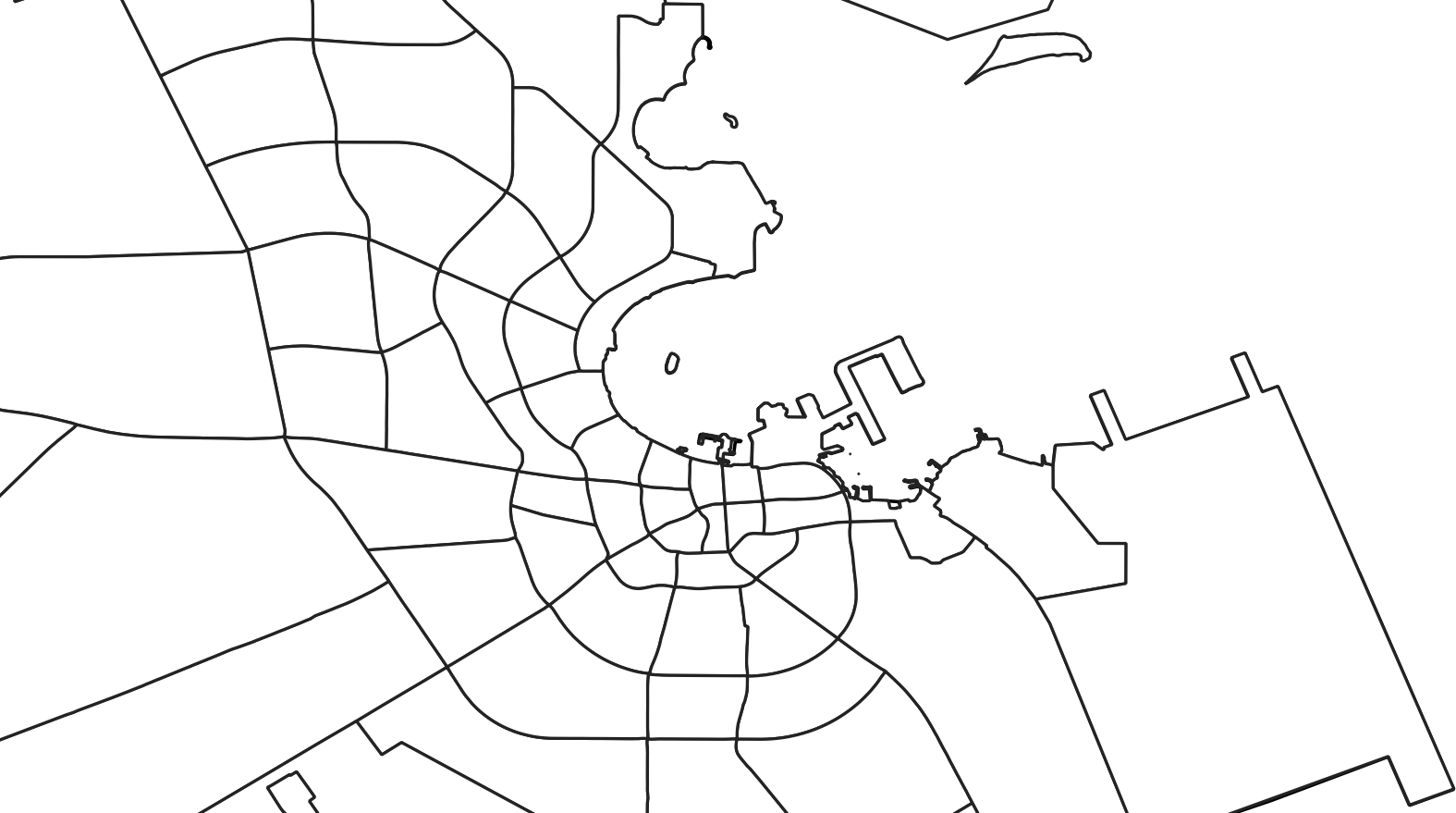}}
\subfigure[Transportation zones]{\label{fig:dense_doha}\includegraphics[width=0.49\columnwidth]{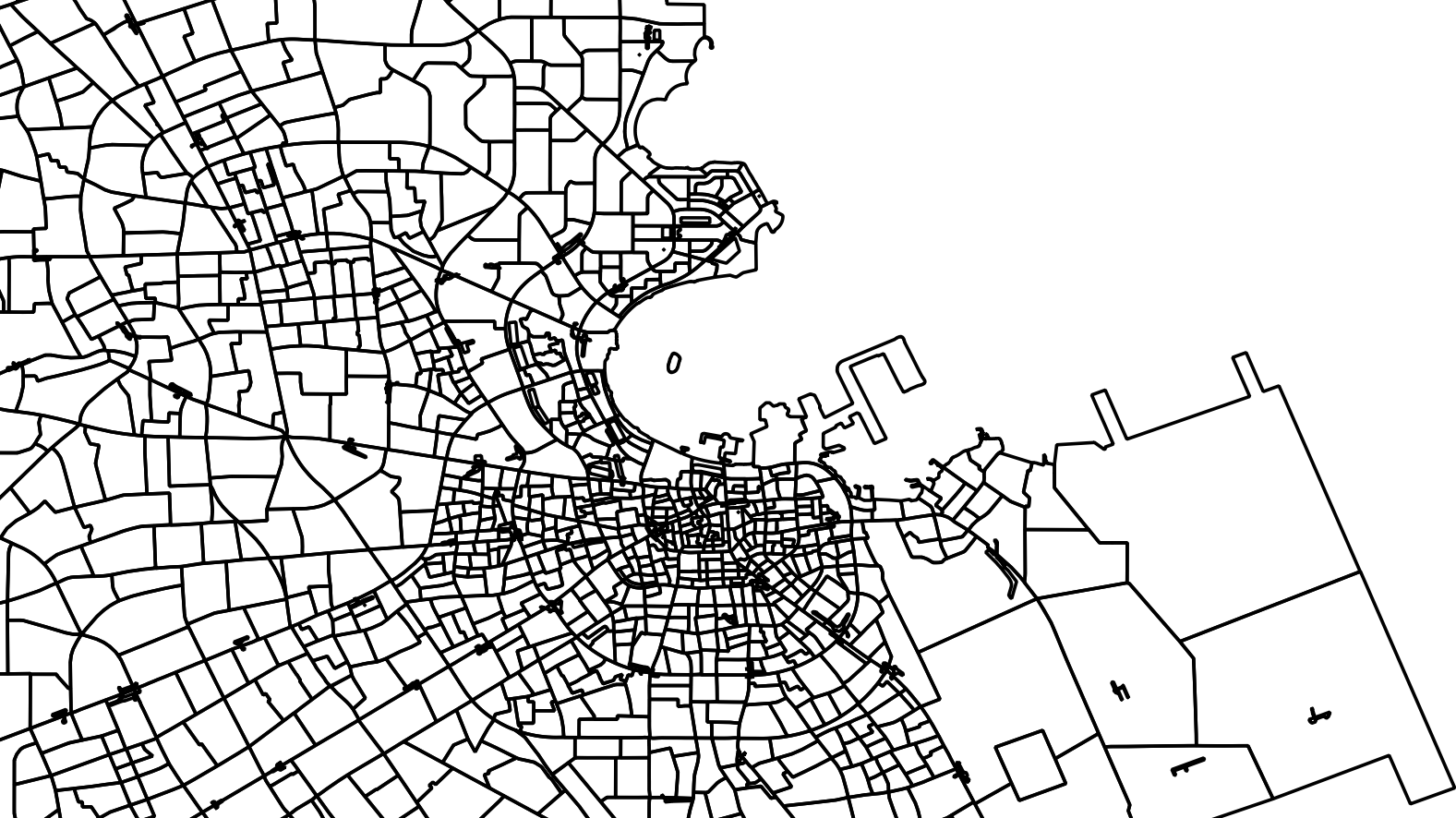}}
\caption{Example of administrative vs. transportation zones in Doha} 
\label{fig:zones}      
\end{figure}

Intuitively, in the free-flow scenario where there is no traffic at all, the travel time of a trip is governed by two parameters: the length of the trip and the maximum speed allowed on different road segments. With traffic, there are two more parameters that impact the travel time: origin and destination locations, and departure time. In other words, the overhead caused by traffic on the free flow travel time varies significantly from an area to another, from an hour to another, from a day to another. Thus, the need to split space and time into smaller units that allow capturing the varying impact of traffic (overhead) in time and space.  
There are different way the space, i.e., a city in our case, can be partitioned. One common practice consists in dividing the space into a grid of equal-sized squares, e.g., $100m \times 100m$. This technique has two major drawbacks: First, it destruct the existing road network boundaries by crossing roads and junctions at random locations. Second, it does not account for spatial population and road distributions, yielding zones with no roads or no trips. Hence, we opt for the use of administrative boundaries to split the space into zones. This practice is very common in transportation research, and such data is abundantly available online\footnote{https://www.geofabrik.de/data/}. Administrative zones are usually generated in a way that preserves the natural road network boundaries and population distribution. It is worth mentioning that there are more specialized partitioning schemes such as transportation area zones (TAZ), which are more adequate for transportation related studies. However, given that TAZ are not always publicly available, we use administrative boundaries in this paper and evaluate the impact of different partitioning schemes in Section~\ref{sec:impact_zone}. Figure~\ref{fig:zones} shows samples of zones in Doha with different granularities. Once the space is partitioned, it is easy to assign trip's origin and destination locations to specific zones.


Similarly to space, time is also often partitioned into small intervals, called time windows. The size of the intervals is typically between 15 mins to 1 hour, depending on the amount and coverage of the trip data available. The finer the granularity, the higher the sparsity. In hour case, we assign each trip $\Gamma_i$ to the time window corresponding to its starting time. We also derive a set of temporal features as shown in the next section. 

\subsection{Feature engineering}
Now, we enumerate the different features that \texttt{STAD} uses to learn good spatio-temporal adjustments of travel times. Some of these features are spatial, others are temporal, and some of them are related to historical traffic conditions that capture recurrent dynamics such as morning and evening commutes in different zones.  

\begin{itemize}
    \item{\it $z_o$ -- origin zone }: this is a unique "alphanumeric" identifier of the zone into which falls the origin location of the trip. Inferred from $o=(lon^o,lat^o)$.
    \item{\it $z_d$ -- destination zone }: identifier of the destination zone. Inferred from $d=(lon^d,lat^d)$.
    \item {\it $h_d$ -- hour of the day}: this is an integer in \{0, 2, \ldots 23\} representing the hour of the day in local time; 00 is for midnight.
    \item {\it $d_w$ -- day of the week}: an integer in \{1, 2, \ldots 7\}; 1 is for Monday and 7 is for Sunday.
    \item {\it $h_w$ -- hour of the week}: an integer in \{0, 2, \ldots 167\} that captures the hours of the week (24x7).
    \item{\it $\tau^{ff}$ -- free-flow travel time}: the travel time returned by \texttt{RE} for a given pair $o,d$. Expressed in seconds.
    \item{\it $l_d$ -- driving distance}: the length of the path returned by \texttt{RE} for the pair $o,d$. Expressed in meters.
    \item{\it $l_h$ -- Haversine distance}: this is the straight line distance between ($o,d$). Expressed in meters.
    \item {\it $p_c$ -- path complexity}: this is the ratio between the $l_d$ and $l_h$; $p_c = l_h/l_d$. The closest the value to one, the simplest is the driving route between $o$ and $d$.
    \item {\it $tr_{z_o}^{*}$ -- trips departing from $z_o$}: ratio of trips departing from $z_o$ to all trips. This captures how busy are the outgoing roads from the orign zone, $z_o$.
    \item {\it $tr^{z_d}_*$ -- trips arriving at $z_d$}: ratio of trips arriving to $z_d$ to all trips. Captures the business of incoming roads to the destination zone, $z_d$. 
\end{itemize}

\subsection{Model Selection}
For each trip $\Gamma_i=(o_i, d_i,t_i,\tau_i,tr^i,p_i) \in \mathcal{T}$, we build a vector feature $x_i=<z_o, z_d, h_d, d_w, h_w, \tau^{ff}, l \ldots, tr^{z_{d}^i}_{*}>$ using all features defined in the previous section. Next, we create for each $\Gamma_i$ a pair $(x_i, \tau_i)$ where $x_i$ is the feature vector and $\tau_i$ is the actual travel time reported for $\Gamma_i$. The objective now is to find a function $F$ that maps $x_i$ to $\tau_i$, such that the following quantity is minimised: 
\[
\frac{1}{|\mathcal{T}|} \times \sum_{tr_i\in\mathcal{T}} L(\tau_i, F(x_i)),
\]

where $L$ is a loss function that captures the errors between exact travel times and adjusted ones. 
This is a classical supervised machine learning framework, regression to be more specific, where one can try different algorithms and pick the best achieving one. However, given that our features are a mix of numerical and categorical, we opted for the tree-based ensemble method called Gradient Boosting \cite{friedman2001greedy}. We know that categorical features can be converted into binary numerical features that work with more algorithms such as SVM and linear regression, however, when the number of categories is high (which is the case of number of zones for instance), this process results in very sparse feature vectors with high dimensionality which requires non trivial treatment via regularizers.    
We set the loss function $L$ to be least squares and fine-tune the Gradient Boosting Regressor to find the best combination of hyper parameters, i.e., number of trees and max depth of each tree. At the end of this step, we obtain a model $\mathcal{M}$ that is capable of predicting accurate traffic-aware travel time estimates $\hat{\tau}$ for any feature vector $x_i$. 

\subsection{Online querying}
The online step of \texttt{STAD} is quite simple and efficient (See algorithm~\ref{alg:online}). This is important not to cause any overhead to the already optimized routing engine which can be deployed in production, i.e., serves thousands of routing requests per second. 
Given a trip request $q=<o, d, t^d>$ with origin, destination, and departure time, and a prediction model $\mathcal{M}$, we first create the feature vector by mapping locations to zones (lines 2--3), extracting temporal features from departure time (line 4), and querying the basic routing engine for free-flow travel-time estimates and driving distance. Next, we compound all these features into one vector and run the model method (line 6.) Other features defined in the previous sections can easily be added here. 

\begin{algorithm}
\caption{Online phase of \texttt{STAD}}
\label{alg:online}
\begin{algorithmic}[1]
\small
    \STATE {\bf Input:} $\mathcal{M}$ - ML predictive model trained offline. $q=<o, d, t^d>$ - a trip request 
    \STATE $z_o = locate(o)$; \COMMENT{find area zone of origin}
    \STATE $s_d = locate(d)$; \COMMENT{find area zone of destination}
    \STATE $h_d, h_w, d_w = extract\_temporal\_features(t^d)$
    \STATE $\tau^{FF}, l = RE(o,d)$; \COMMENT{generate basic free-flow information for $q$}
    \STATE $\hat{\tau} = \mathcal{M}.predict(z_o, z_d, h_d, h_w, d_w, \tau^{FF}, l)$
    \RETURN $\hat{\tau}$
\end{algorithmic}
\end{algorithm}


\section{\texttt{STAD$^*$}: Live Traffic Integration}
\label{sec:stad_star}
While \texttt{STAD} is good at capturing recurrent traffic patterns such as morning and evening commutes, and weekends vs. weekdays, it is not well suited to deal with unexpected events such as sudden congestion, accidents, processions, or severe weather conditions such as heavy rain. 
Thus, in the case live traffic is available us, we devise \texttt{STAD$^*$}, a variant of \texttt{STAD} capable of integrating (near) live traffic conditions into the adjustment process of travel time estimations. 

One simple way to take into account live traffic information is to compare the current traffic status to the expected one from historical data. This can be achieved, by monitoring the average speed of vehicles. We propose to monitor the traffic level via a congestion index $\mathcal{CI}$. 
Assuming that time is partitioned into windows, i.e., $\mathcal{TW} = \{tw_0 \ldots tw_{24}\}$ to capture hourly traffic patterns for a typical day, we could easily compute offline the average speed observed at each time window. At query time, assuming that we would like our system to update to live traffic every $10 mins$, we compute the average speed of all trips ($\mathcal{T}_c$) completed in the last $10 mins$ and compare it to the expected speed from the time window that corresponds to the last 10 minutes.  
Equation (1) gives the formula used to compute $\mathcal{CI}$. A value greater than 1.0 indicates a traffic that is less congested than expected whereas a value lower than 1.0 indicates a traffic level that is more congested than expected. 

\begin{equation}
\mathcal{CI}_{tw} = \frac{\frac{1}{|\mathcal{T}_c|} \sum_{\Gamma_i \in \mathcal{T}_c} l_i/\tau_i}{\frac{1}{|\mathcal{T}_{tw}|} \sum_{\Gamma_j \in \mathcal{T}_{tw}} l_j/\tau_j}
\end{equation}

$\mathcal{CI}$ makes it easy for instance to distinguish an 8am of a busy Monday from the fluid 8am of a quite Sunday. Even in cases where time is partitioned into finer windows (e.g., $24 \times 7 = 168$) to capture weekly patterns, $\mathcal{CI}$ allows to distinguish a 10am of a busy Tuesday from a 10am of calm Tuesday that coincides with a school break for instance.   
Depending on the values of $\mathcal{CI}$, \texttt{STAD}$^*$ will learn to increase or decrease the adjusted travel time $\hat{\tau}$ to reflect the current traffic condition. The following additional features can be derived from Equation (1):

\begin{itemize}
    \item{\it $\mathcal{CI}_{h_d}$ -- global congestion index - day}: Congestion index by comparing current speeds to average speeds of the hour of the day ($h_d$) to which the departure time of the request falls into.
    \item{\it $\mathcal{CI}_{h_w}$ -- global congestion index - week}: Congestion index by comparing current speeds to average speeds of the hour of the week ($h_w$) to which the departure time of the request falls into.
\end{itemize}

These two features are added to the ones defined in the previous section in order to train a new predictive model $\mathcal{M}*$ for \texttt{STAD}$^*$. 



\section{Experiments}
\label{sec:experiments}

In this section, we describe our evaluation setup, algorithms and baselines, metrics, data we used for different experimental evaluations.
 
\subsection{Algorithms and Baselines} 
In order to assess the importance of different features on the accuracy achieved by our system, \texttt{STAD}, we evaluated four different variants: \texttt{STAD}$_t$ which only uses temporal features (i.e.,$\tau^{ff}$, $l_d$, $h_d$, $d_w$, and $h_w$), \texttt{STAD}$_{st}$ which uses spatio-temporal features (i.e., $\tau^{ff}$, $l_d$, $h_d$, $d_w$, $h_w$, $z_o$, and $z_d$), \texttt{STAD}$_{all}$ which uses all features described in Section~\ref{sec:stad}, and finally \texttt{STAD}$^*$ that integrates live traffic data. Without loss of generality, \texttt{STAD} and \texttt{STAD}$_{st}$ are used interchangeably in the remaining of this section. We compare our systems to three different baselines:

\subsubsection{\texttt{RE}}
We use OSRM\cite{luxen-vetter-2011} as our default traffic oblivious routing engine (\texttt{RE}). OSRM uses the road network extracted from OpenStreetMaps\footnote{https://www.openstreetmap.org} to build its internal graph. The system relies on some prior knowledge of traffic in order to compute free-flow weights to each road segment. These weights are generated based o metadata associated with road segments, including type of road (e.g., highway), posted speed (e.g., 50kmph), and penalty scores for different types of turns (e.g., left-turn, U-turn).

\subsubsection{\texttt{KNN}}
In addition to \texttt{RE}, we implemented \texttt{KNN}, a nearest neighbors based algorithm developed by Wang et al. ~\cite{wang2016simple, wang2019simple}. Similar to \texttt{STAD}, \texttt{KNN} combines spatio-temporal aspects of trips to predict travel time of any trip query $q=(o_q, d_q, t_q)$. The algorithm assumes a collection of historical trips $\mathcal{T}$ each of which is defined as 5-tuple $\Gamma_i = <o_i, d_i, t_i, l_i, \tau_i>$ where $o_i,d_i$ are origin and destination locations, $\tau_i, l_i$ are respectively the travel time and the distance of the trip, and  $t_i$ is the departure time. Given a query $q$, \texttt{KNN} computes the expected travel time of $q$ as follows:
\[
\hat{q} = \frac{1}{|\mathcal{N}(q)|} \sum_{\Gamma_i \in \mathcal{N}(q) } \tau_i \times \frac{V(t_i)}{V(t_q)},
\]

where $\mathcal{N}(q)$ is the set of nearest neighbor trips to $q$. A trip $\Gamma_i$ is considered as a nearest neighbor to $q$ if their origins and destinations are close enough. In our adaptation, we assume two trips as nearest neighbors if their origins and destinations fall into the same spatial zones respectively. \texttt{KNN} splits time into 168 windows corresponding to the hours of a typical week, then $V(t_i)$ (resp. $V(t_q)$ is the average speed observed for all trips that started at the hour corresponding to the departure time $t_i$ (resp. $t_q$ departure time of $q$.) Speeds can easily be inferred from the distance $l_i$ and travel time $\tau_i$ of any trip $\Gamma_i$

\subsubsection{\texttt{Google} Maps} We also compare our findings against one of the leading commercial map engines in the market, Google Maps. Using their Direction API~\footnote{https://developers.google.com/maps/documentation/directions/start}, Google Maps allows developers to query for best routes for any request $q=(o_q, d_q, t_q)$ such that $t_q$ is sometimes in the future. Google Maps uses then historical traffic data to compute the travel times of these requests. We sampled uniformly at random 2K different trip requests that we submitted to Google Maps. Since the timestamp need to be in the future we submitted our queries in the last week of January with the timestamps falling in the second week of February. For a fair comparison, timestamps used in Google maps queries correspond to the same hour/weekday/month of each trip in our dataset. An important remark here is that even though our Doha dataset is about 2 years old, the road network as well as the traffic pattern have experienced a nontrivial change between 2018 and 2020, albeit quantifying that change is difficult and out of scope of the present paper. The data in NYC and Porto are from 2013 and we believe that traffic patterns over 6-7 years change substantially, and do not compare these 2 cities with Google Maps API in this paper.

\subsection{Evaluation metrics}
In order to assess the accuracy and quality of travel time estimates produced by different algorithms, we use mean absolute error (MAE) and median absolute error (MedAE), which have both seconds as units. However, because the values of these metrics depend on the travel time distribution in our datasets, it important to report relative errors as well. Thus, we compute the mean absolute percentage error (MAPE) also knows as mean relative error as well as the median absolute percentage error (MedAPE.)  Formulas for the different metrics are defined as follows: 
$MAE(Y, \hat{Y}) = 1/n \sum_{i=1,n}|y_i-\hat{y_i}|$, $MedAE(Y, \hat{Y}) = median(\{|y_i-\hat{y_i}|\}_{i=1,n})$, $MAPE(Y, \hat{Y}) = 100 \times 1/n \sum_{i=1,n}|y_i-\hat{y_i}|/y_i$, and finally $MedAPE(Y, \hat{Y}) = 100 \times median(\{|y_i-\hat{y_i}|/y_i\}_{i=1,n})$

\subsection{Dataset description}
\label{sec:data}
\begin{figure}[h]
\centering
\includegraphics[width=0.7\columnwidth]{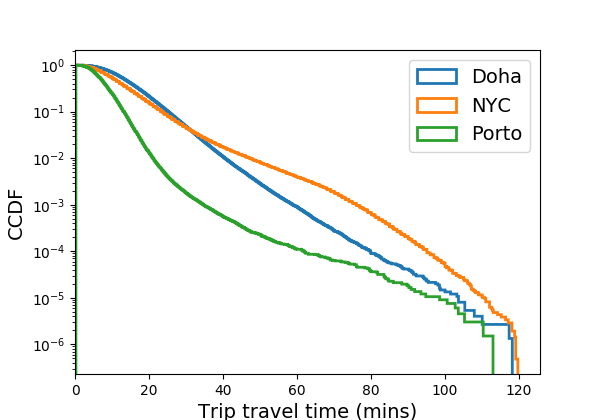}
\caption{Complementary cumulative distribution function of travel times estimates for Doha, NYC, and Porto (y-axis is in log scale.)}
\label{fig:ccdf}
\end{figure}

We use trips data generated by Taxi companies operating in three different cities: Doha (Qatar), NYC (US), and Porto (Portugal). 
Table~\ref{tbl:datasets} presents some statistics about the datasets. The time span covered by trips varies from 1 week in 2013 for NYC, to 5 weeks in 2018 for Doha, to 1 year 2013/2014 for Porto. We also found a significant discrepancy in the distribution of travel times among the three cities. Doha for instance has a median trip duration of 13.26 minutes compared to 10.55 minutes in NYC and 7 minutes in Porto. Complementary cumulative distribution functions of travel times are illustrated in Figure~\ref{fig:ccdf} which shows that around 90\% of trips last less than 12 minutes in Porto (rapid decay) versus 30 mins for Doha and NYC. 

\begin{table}[ht]
\caption{Characteristics of different datasets used in the experimental evaluation.}
\begin{center}
\begin{tabular}{|c||c|c|c|c|}
\hline
\textbf{city} & {\bf dates} & {\bf \#trips} & {\bf med. TTE}& {\bf trips/day} \\ 
\hline
\hline
Doha & 2018/01/01-2018/02/08 & 748,096  & 13.26 & 19.2K \\
\hline
NYC & 2013/11/01-2013/11/07  & 2,078,503  & 10.55 & 297K \\
\hline
Porto & 2013/07/01-2014/07/01  & 662,614 & 7.0 & 2.5K \\
\hline
\end{tabular}
\end{center}
\label{tbl:datasets}
\end{table}

\subsection{Main results}
We train different \texttt{STAD} models on 70\% of the trips and test on the remaining 30\%. Reported results are averages over multiple runs.

\begin{figure*}[ht]
\centering
\subfigure[Doha]{\label{fig:doha_maes}\includegraphics[width=0.32\linewidth]{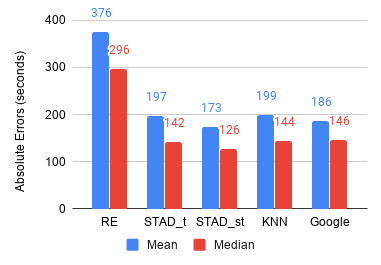}}
\subfigure[NYC]{\label{fig:nyc_maes}\includegraphics[width=0.32\linewidth]{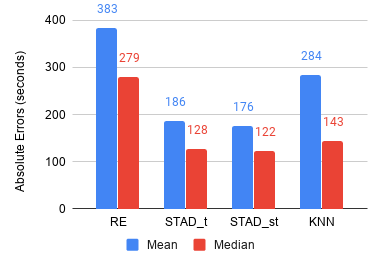}}
\subfigure[Porto]{\label{fig:porto_maes}\includegraphics[width=0.32\linewidth]{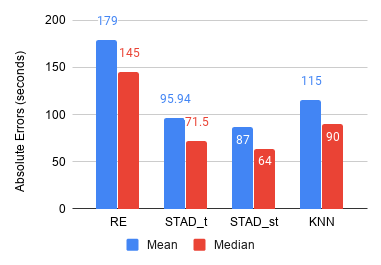}}

\subfigure[Doha]{\label{fig:doha_mapes}\includegraphics[width=0.32\linewidth]{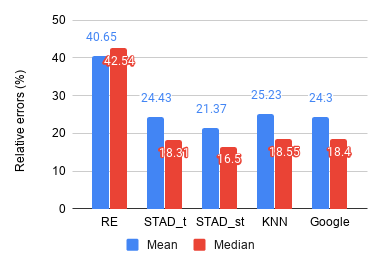}}
\subfigure[NYC]{\label{fig:nyc_mapes}\includegraphics[width=0.32\linewidth]{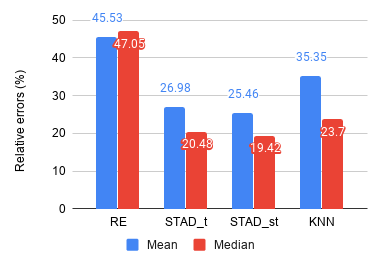}}
\subfigure[Porto]{\label{fig:porto_mapes}\includegraphics[width=0.32\linewidth]{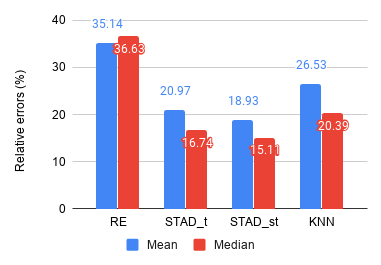}}

\subfigure[Doha]{\label{fig:doha_mapes_hourly}\includegraphics[width=0.32\linewidth]{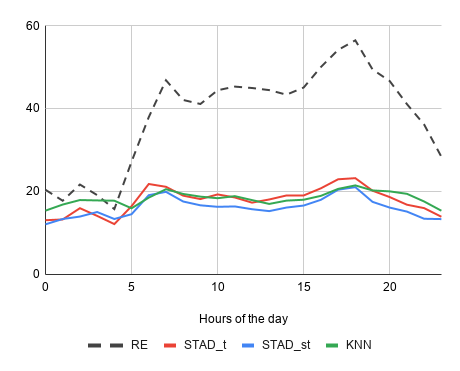}}
\subfigure[NYC]{\label{fig:nyc_mapes_hourly}\includegraphics[width=0.32\linewidth]{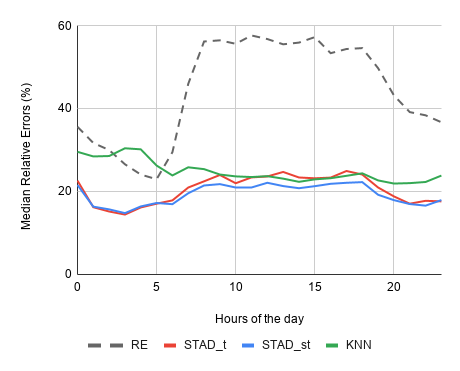}}
\subfigure[Porto]{\label{fig:porto_mapes_hourly}\includegraphics[width=0.32\linewidth]{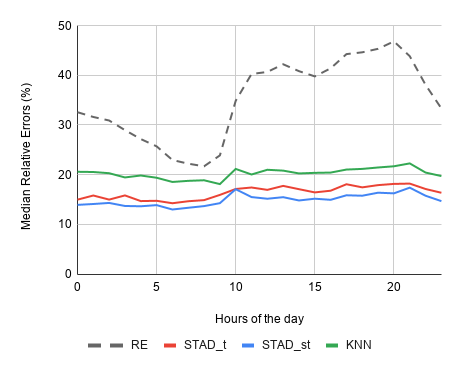}}
\caption{Main results for travel time predictions. The first row reports mean and median absolute errors of travel time estimations for unseen trips in three different cities. The second row reports mean and median relative errors in percentages for the same trips. The third row reports hourly median relative errors.} 
\label{fig:main_res}      
\end{figure*}

Figure~\ref{fig:main_res} illustrates the main results for travel time estimation using the three cities. In the first row, we report the mean and median absolute errors achieved by different algorithms in predicting travel times. The second row reports global relative errors whereas the third one reports hourly relative errors. The foremost observation to be made here is that \texttt{STAD}$_{st}$ outperforms by far all other algorithms in all three cities and in all considered metrics. Second, we clearly see that accounting for spatial features (\texttt{STAD}$_{st}$) yields significant improvement compared to \texttt{STAD}$_t$ which uses temporal features only. This is shown in Figures ~\ref{fig:doha_maes},\ref{fig:nyc_maes}, and \ref{fig:porto_maes} which report respectively a reduction in median absolute errors induced by the spatial features of 16 seconds in Doha, 6 seconds in NYC, and 7.1 seconds in Porto. The gain in relative errors (second row of Figure~\ref{fig:main_res}) is of two points in Doha (18.31\% to 16.5\%), 1.5\% in Porto and 1\% in NYC. 

To our surprise, the simple \texttt{STAD}$_t$ did better than \texttt{KNN} in all experimental configurations, let alone \texttt{STAD}$_{st}$. In Doha for instance, \texttt{KNN} achieves a median average error of 144 seconds vs. 142 seconds for \texttt{STAD}$_t$ and a low 126 seconds for \texttt{STAD}$_{st}$. The absolute errors of \texttt{KNN} are also 15\% higher than \texttt{STAD}$_{st}$ in NYC and 29\% higher in Porto. In terms of relative errors, \texttt{STAD}$_{st}$ is 2 points lower than \texttt{KNN} in Doha (16.5\% vs. 18.5\%), 4 points lower in NYC (19.42\% vs. 23.7\%), and 5 points lower in Porto (15.11\% vs. 20.39\%). These big improvements of \texttt{STAD}$_{st}$ over \texttt{KNN} are intriguing when we know that the latter is designed to embed the spatio-temporal aspects. Our interpretation is that relying on origin and destination to declare two trips similar is not a good idea. One could at least account for trips distance to remove outliers. Also, simply averaging travel times of similar trips is not good enough, especially when all nearest neighbor trips are considered regardless of the time at which they were made.

As expected, the experiments show that even for a state-of-the-art routing engine which incorporates a significant prior knowledge about traffic such as delays at junctions, traffic lights, turn penalties, it is very difficult to get travel time estimates right without traffic information. This is reflected in the high median absolute percentage (relative) errors achieved by \texttt{RE} in Doha (42.54\%), NYC (47.05\%), and Porto (36.63\%) as shown in Figures~\ref{fig:doha_mapes},\ref{fig:nyc_mapes}, and \ref{fig:porto_mapes}. However, it is interesting to note that free-flow travel time estimates produced by \texttt{RE} are quite comparable to other methods for specific time windows (00:00 -04:00am in Doha, with a MedAPE=20\%.) In the case of NYC, \texttt{RE} outperforms \texttt{KNN} between 03:00-05:00am. This is worth exploring for companies who want to cut some of the cost that goes to commercial maps.

As planed, we also tried Google maps services for Doha and reported results in Figures~\ref{fig:doha_maes} (absolute errors) and \ref{fig:doha_mapes} (relative errors.) In terms of relative errors, Google is slightly better than \texttt{KNN} with a median percentage error of 18.4\% compared to 18.55\%. However, both version of \texttt{STAD} are better than Google with a clear win of \texttt{STAD}$_{st}$ which achieves a median percentage error of 16.5\%. Similarly, the median absolute error achieved by Google is 146 seconds which is quite similar to \texttt{KNN} with 144 seconds and \texttt{STAD}$_t$ with 142 seconds. But once again, \texttt{STAD}$_{st}$ shows a clear superiority with 126 seconds of median percentage error. We deliberately decided not to compare with NYC and Port trips because they are more than 7 years old which makes it unfair comparison. 

We report in the third row of Figure~\ref{fig:main_res} the hourly median percentage errors of different algorithms in different cities. Each curve has 24 points corresponding to the 24 hours of the day. In the case of Doha for instance, we clearly distinguish two error spikes corresponding to morning (06:00am-07:00am) and evening (05:00pm-06:pm) commutes in which the percentage error is above 20\% whereas the errors are below the bar of 20\% for all other hours of the day. This behavior is partly due to the fact that traffic is more unpredictable during rush hours, when incidents tend to happen more often. In the remaining hours of a typical day, where traffic shows more stability and seasonality, all algorithms seem to do well, with a net superiority for \texttt{STAD}$_{st}$ though. The same kind of pattern is observable in NYC (Figure~\ref{fig:nyc_mapes_hourly}) and Porto (Figure~\ref{fig:porto_mapes_hourly}), thought for the particular case of NYC the error pattern looks two-phased: day (07:00am to 07:00pm) where errors are above 20\% and night (07:00pm to 07:00am) where errors go below 20\%. This might be explained by the constant business of the city throughout the day.

Finally, we report in Table~\ref{tbl:stads} results achieved by different version of \texttt{STAD}. We clearly see that accounting for spatial features yields significant improvements, i.e median absolute error of \texttt{STAD}$_{st}$(126 secs) is 10\% better compared to \texttt{STAD}$_t$ (141 secs). However, adding more features ~\texttt{STAD}$_{all}$ (125 secs) yields only 1\% improvements over \texttt{STAD}$_{st}$.

\begin{table}
\caption{Results of different \texttt{STAD} variants on Doha data.}
\begin{center}
\begin{tabular}{|c||c|c|c|c|}
\hline
\textbf{Variant} & {\bf MAE(sec)} & {\bf MedAE(sec)} & {\bf MAPE(\%)}& {\bf MedAPE(\%)} \\ 
\hline
\hline
\texttt{STADt}$_t$ & 196.78 &	141.43	& 24.35 & 18.4 \\
\hline
\texttt{STAD}$_{st}$ & 174.49	& 126.76 & 21.40 & 16.50 \\
\hline
\texttt{STAD}$_{all}$ & 172.14 & 125.42 & 21.01 & 16.33 \\
\hline
\end{tabular}
\end{center}
\label{tbl:stads}
\end{table}

\subsection{Impact of number of trips}
\begin{figure}
\centering
\subfigure[Median Absolute Errors (sec.)]{\label{fig:size_eff_medaes}\includegraphics[width=0.49\columnwidth]{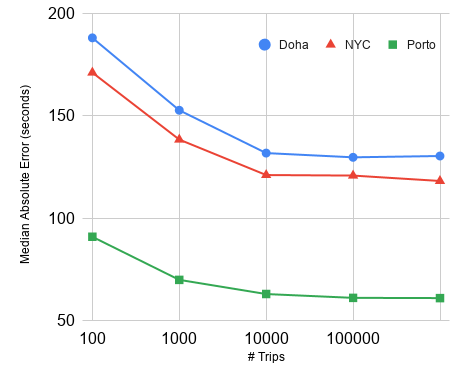}}
\subfigure[Median Relative Errors (\%)]{\label{fig:size_eff_medapes}\includegraphics[width=0.49\columnwidth]{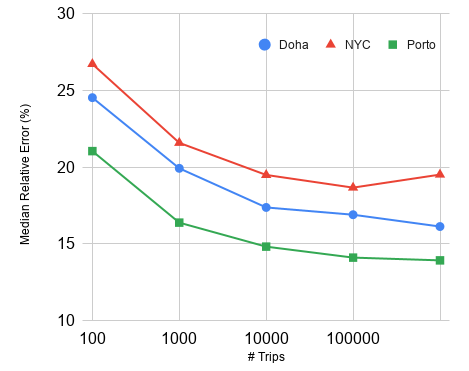}}
\caption{The effect of number of trips available for training on the accuracy of travel time predictions of STAD\_st in the three cities.}      
\label{fig:size}
\end{figure}

We investigate the impact of the number of trips available for training \texttt{STAD} on the accuracy of the predicted travel times in all cities. 
%
Figures~\ref{fig:size_eff_medaes} and \ref{fig:size_eff_medapes} report respectively median absolute and relative errors in the three cities, function of the number of trips used in the training. The elbow shaped curves we see indicate a diminishing return property according to which the impact of adding more data reduces as the size of data increases. For instance, we observe that absolute errors drastically improve with more trips until we reach the breaking point of 10K trips. Adding more trips beyond 10K does not seem to yield any significant reduction in terms of absolute error for all three cities. However, if the objective is to optimize for median percentage errors, then it seems that 100K is the magic number of trips needed. In the case of NYC, we even see an inverse phenomenon where the percentage error increases when more trips are used, which can be due to over-fitting. 

\subsection{Impact of real-time traffic data}
When live traffic is available i.e., the possibility to access trips as soon as they are completed, it becomes possible to deploy \texttt{STAD}$^*$ which incorporates congestion indices into the adjustment model. Table~\ref{tbl:stad_star} shows comparative results between \texttt{STAD} (historical traffic model) and \texttt{STAD}$^*$. As expected, live traffic information yields significant improvements in the accuracy of travel time of about 5 seconds in median absolute error ($\approx 4\%$ reduction). Mean absolute percentage error gains almost 1 point lowering from 21.27\% for \texttt{STAD} to 20.38\% for \texttt{STAD}$^*$.    

\begin{table}[h]
\caption{The impact of real-time traffic data on travel time estimation accuracy in Doha dataset.}
\begin{center}
\begin{tabular}{|c||c|c|c|c|}
\hline
\textbf{Variant} & {\bf MAE(sec)} & {\bf MedAE(sec)} & {\bf MAPE(\%)}& {\bf MedAPE(\%)} \\ 
\hline
\hline
\texttt{STAD} & 161.11 & 122.30	& 21.27 & 16.20 \\
\hline
\texttt{STAD}$^*$ & 158.31 & 117.37 & 20.38 & 15.75 \\
\hline
\end{tabular}
\end{center}
\label{tbl:stad_star}
\end{table}

As mentioned earlier, the real advantage of \texttt{STAD}$^*$ over \texttt{STAD} is its ability to correct for specific events and for days that are substantially different from those seen in training on historical data. For instance, we found that \texttt{STAD}$^*$ yields to significant reduction in median absolute error for trips happening on Fridays, especially for those taking place during time intervals considered as rush hours in other days (Note that Friday is a weekend in Doha and its traffic patterns are quite similar to those observed on Sunday in the west.) That is, \texttt{STAD}$^*$ saves 21 seconds for trips starting at 5pm, 11 seconds for those starting at 6 pm (evening commute window). Likewise, we observe a reduction of 13 seconds for 7am and 12 seconds for 6am (morning commute.) 
It is worth mentioning that the evaluation setup for this comparison is different from the one we used to produce results of Table~\ref{tbl:stads}, which explains the differences in scores obtained by \texttt{STAD\_st} in the two tables. Indeed, to account for live traffic, we needed to sort trips by departure time. We removed then the first 200K trips, used to learn historical average speeds for different time windows, which represents around 26\% of all trips in Doha dataset. 

\subsection{Impact of spatial partitioning choice}
\label{sec:impact_zone}
\begin{figure}
\centering
\subfigure[Absolute Errors (sec.)]{\label{fig:taz_medaes}\includegraphics[width=0.49\columnwidth]{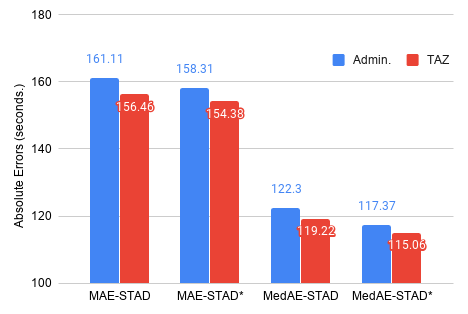}}
\subfigure[Relative Errors (\%)]{\label{fig:taz_medapes}\includegraphics[width=0.49\columnwidth]{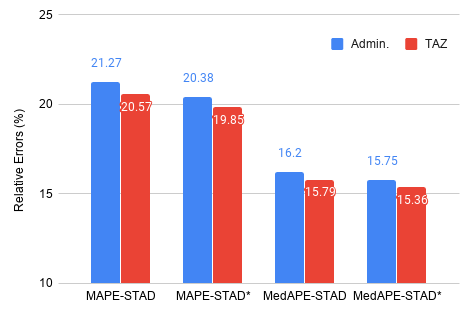}}
\caption{Comparative results for different spatial partitioning schemes on accuracy of travel times in Doha.}
\label{fig:taz}
\end{figure}

We investigate now the impact of way the space is partitioned on the accuracy of travel time estimation. We have deliberately decided to use administrative zoning as a default choice for partitioning, which can be easily obtained for almost all cities around the world. However, it is well known that transportation departments around the world have their own way of partitioning a city into geography units knows as traffic/transportation analysis zones (TAZ), which takes into account population densities as well as the geometry of the road network. Thanks to our collaboration with Ministry of Transport and Communication in Qatar, we could obtain their TAZ shapefile which contains 1,839 zones. We re-assigned origin and destination locations of every trip into one of these zones, and re-run both \texttt{STAD}$^*$ and \texttt{STAD} on the new data.
We show in Figure~\ref{fig:taz} comparative results achieved using two different spatial partitioning schemes. Overall, we notice that TAZ yields better accuracy compared to regular administrative zones. The gain in median (resp. mean) absolute error is of 3 secs (resp. 5 secs) for \texttt{STAD} and 2 seconds (resp. 4 secs) for \texttt{STAD}$^*$. The gain in relative errors is less significant though. 



\section{Related Work}
\label{sec:related}
The abundance of transportation data in the form of GPS trajectories and trips has opened new horizons for capturing traffic dynamics in road networks. We can distinguish three main categories of work aiming at estimating trips travel time: segment-based, path-based, and origin-destination based. 

{\it Segment-based.} Also known as link travel time, is the most common approach in which GPS data (trips or trajectories) are used to learn the weights of individual edges of the road graph. Travel time of a path is then calculated as the sum of weights on its constituent edges~\cite{ide2009travel,yuan2011t,ide2011trajectory,zheng2013time,wang2014travel,aljubayrin2016finding,wedge18}. Several techniques have been explored to efficiently and accurately learn these weights. For instance, \cite{ide2009travel,ide2011trajectory,zheng2013time, wedge18, mapreuse19} use different types of regression (e.g., linear, ridge, Gaussian), some times with regularizers, to effectively learn the weights. Similarly, \cite{yuan2011driving} uses Markov chains to infer future travel times, whereas \cite{yang2013travel} proposes to use a spatio-temporal extension of Hidden Markov models.   
In order to deal with the inherent data sparsity problem in which not all segments can be covered with data, several authors have adopted matrix decomposition techniques for missing values completion and travel time prediction~\cite{deng2016latent, yu2016temporal, zahma19}. In most of the cases, the decomposition is done with the introduction of either some latent features for the road network (e.g., \cite{deng2016latent}) or some particular temporal regularizers (e.g., \cite{yu2016temporal}.)

{\it Path-based.} One of the main shortcomings of segment-based techniques is their failure to capture inter-link transitions such as turns at intersections or waiting at traffic lights. Some approaches tackled this problem by explicitly estimating the time spent at junctions ~\cite{herring2010using, li2015inferring}. Another line of work aims at concatenating segments into (sub-)paths and estimating travel time directly for paths instead of individual segments~\cite{wang2014travel,aljubayrin2016finding,dai2016path,yang2018pace}. In ~\cite{wang2014travel}, authors propose to use a three dimensional tensor (users, segments, time) combined with frequent trajectory patterns to find the best combination of sub-paths to use in online travel time estimation. The solution is shown to find a good trade-off between the length of sub-paths and number of trajectories traversing them. Dai et al. ~\cite{dai2016path} take a different approach to the same problem and introduce the concept of hybrid graphs to accurately capture the cost distribution of paths. The idea is to learn a weighting function for paths instead of edges, then find a good approach to combine distributions of multiple sub-paths to estimate the travel time of a query path. A follow-up to this work is PACE ~\cite{yang2018pace} in which authors solve the problem in a more principled way by tackling two problems: first, estimation of path cost distributions using an optimal set of trajectories; second, finding the rights paths for a source-destination pair. 

{\it OD-based.} While path-based techniques overcome the inter-link transition problem, they do introduce some difficult challenges such as finding the right sub-paths for a query trip, which is often solved in an ad-hoc mode or using heuristics.
More recently, some authors looked at inferring travel time estimates for pairs of origin-destination without computing paths~\cite{wang2016simple,jindal2017unified,hu2018recurrent,li2018multi,wang2019simple}. This is particularly relevant in online configurations where path information is not available or costly to obtain. Examples of applications where this is needed include taxi dispatching where there is a huge need to compute travel time estimates between locations of a set of vehicles and that of a customer. 
One of the earliest works in this space is done by Wang et al.~\cite{wang2016simple, wang2019simple}. Here, authors propose to compute the travel time of a given trip defined with an origin, destination, and departure time, by looking at the travel time of historic nearest neighbor trips. Two trips are assumed to be neighbors if their origins and destinations fall respectively within a predefined radius distance. A scaling factor that captures traffic variations is then computed to allow for a weighted averaging of travel times reported by nearest neighbor trips. A more recent work by Li et al. ~\cite{li2018multi} uses multi-task representation learning for arrival time estimation. The idea is to learn link representations that optimize not only for travel times but also for other related targets such as travel distance and number segments on the path. The authors propose to embed the spatio-temporal aspects of traffic via Laplacian regularization. other "unpublished" yet interesting attempts explored the use of deep neural networks such as GNNs~\cite{hu2018recurrent} and ST-NNs~\cite{jindal2017unified} to travel time estimation tasks.  

\section{Conclusion}
\label{sec:conclusion}
We presented in this paper, \texttt{STAD}, a novel system capable of adjusting free-flow travel time estimates produced by any traffic oblivious routing engine to match actual traffic patterns of a city. \texttt{STAD} partitions the space into areas or zones, and time into windows to capture the spatio-temporal patterns of traffic. We also presented \texttt{STAD}$^*$, a version of our system that adapts the adjustment of travel time to live traffic conditions. Our experiments on real traffic datasets from three cities show that \texttt{STAD} yields significant improvement in terms of travel time accuracy compared to existing methods, or even to premium commercial routing engines such as Google Maps. Indeed, in the case of Doha for instance, our system achieves a median relative percentage error of 16.5\%, whereas KNN based methods achieve 18.55\% and Google Maps 18.40\%. Also, \texttt{STAD}'s median absolute error is 126 seconds compared to 144 second and 146 seconds for KNN and Google Maps respectively.   
In the future, we would like to explore the use of spatially and temporally weighted regression methods to better account the spatio-temporal non stationarity of road traffic. We are also working on releasing an end-to-end open-source system built on top of OSRM to make it easy for people and companies to get started with time-dependent routing and travel time estimation.     

\section*{Acknowledgment}
We would like to thank the Land Transport Planning Department at the Ministry of Transport and Communication in Qatar for their meaningful engagement with our research. We would like also to thank Karwa Technologies at Taxi Mowasalat for their continuous support and fruitful collaboration. 

\bibliographystyle{IEEEtran}
\bibliography{biblio.bib}

\end{document}